\documentclass[journal=jacsat,manuscript=article]{achemso}

\usepackage[version=3]{mhchem} 

\author{Theng-Loo~Lim}
\affiliation{Department of Physics, University of Ottawa, 25 Templeton St, Ottawa, ON, K1N 6N5, Canada}

\author{Yaswant~Vaddi}
\affiliation{Department of Physics, University of Ottawa, 25 Templeton St, Ottawa, ON, K1N 6N5, Canada}

\author{M.~Saad~Bin-Alam}
\affiliation{School of Electrical Engineering and Computer Science, University of Ottawa, 25 Templeton St, Ottawa, ON,  K1N 6N5, Canada}

\author{Lin~Cheng}
\affiliation[North University of China]{School of Instrument and Electronics, North University of China, Taiyuan, 030000, China}

\author{Rasoul~Alaee}
\affiliation{Department of Physics, University of Ottawa, 25 Templeton St, Ottawa, ON, K1N 6N5, Canada}

\author{Jeremy~Upham}
\affiliation{Department of Physics, University of Ottawa, 25 Templeton St, Ottawa, ON, K1N 6N5, Canada}

\author{Mikko~J.~Huttunen}
\affiliation{Photonics Laboratory, Physics Unit, Tampere University, Korkeakoulunkatu 3, Tampere, FI-33014, Finland}

\author{Ksenia~Dolgaleva}
\affiliation{School of Electrical Engineering and Computer Science, University of Ottawa, 25 Templeton St, Ottawa, ON,  K1N 6N5, Canada}

\author{Orad~Reshef}
\affiliation{Department of Physics, University of Ottawa, 25 Templeton St, Ottawa, ON, K1N 6N5, Canada}
\email{orad@reshef.ca}

\author{Robert~W.~Boyd}
\affiliation[University of Ottawa]
{Department of Physics, University of Ottawa, 25 Templeton St, Ottawa, ON, K1N 6N5, Canada}
\alsoaffiliation[University of Rochester]
{Institute of Optics and Department of Physics and Astronomy, University of Rochester, 500 Wilson Blvd., Rochester, New York 14627, USA}

\title{Fourier-Engineered Plasmonic Lattice Resonances}

\begin{document}

\begin{abstract}
Resonances in optical systems are useful for many applications, such as frequency comb generation, optical filtering, and biosensing. However, many of these applications are difficult to implement in optical metasurfaces because traditional approaches for designing multi-resonant nanostructures require significant computational and fabrication efforts. To address this challenge, we introduce the concept of Fourier lattice resonances (FLRs) in which multiple desired resonances can be chosen a priori and used to dictate the metasurface design. Because each resonance is supported by a distinct surface lattice mode, each can have a high quality factor. Here, we experimentally demonstrate several metasurfaces with arbitrarily placed resonances (e.g., at 1310 and 1550~nm) and $Q$-factors as high as 800 in a plasmonic platform. This flexible procedure requires only the computation of a single Fourier transform for its design, and is based on standard lithographic fabrication methods, allowing one to design and fabricate a metasurface to fit any specific, optical-cavity-based application. This work represents an important milestone towards the complete control over the transmission spectrum of a metasurface.

\bf{Key Words}: plasmonics, metasurfaces, lattice resonances, nanoparticle arrays, nanophotonics. 

\end{abstract}

Plasmonic nanoparticles are essential tools for the manipulation of light beams in metasurfaces because of the flexibility of the placement of their localized surface plasmon resonances (LSPRs)\cite{Hutter2004ExploitationResonance}. The LSPR wavelength can be easily tailored since it is highly dependent on the size and the shape of the nanoparticle\cite{maier2007plasmonics}. Additionally, LSPRs confine light in the nanoscale, leading to significant local-field enhancements\cite{Oldenburg1998NanoengineeringResonances}. Thanks to these desirable properties, many applications have been realized in a broad range of research fields such as optical filtering\cite{Yokogawa2012PlasmonicApplications,Zeng2013UltrathinFilters}, harmonic generation\cite{Palomba2009NonlinearAntennas}, and  bio-sensing\cite{Qiu2021Thermoplasmonic-AssistedSARS-CoV-2}. However, the applications of LSPRs are limited by the intrinsic absorption loss of metals; the quality-factor of LSPRs is low ($Q$-factor $<$ 10)\cite{Doiron2019QuantifyingSurvey, Kravets2018PlasmonicApplications}. One method to improve the $Q$-factor of resonant plasmonic systems is to consider instead collective optical responses of multiple nanoparticles in a periodic lattice arrangement; such metasurface-scale responses are known as surface lattice resonances (SLRs) \cite{Utyushev2021CollectiveBeyond, Kravets2018PlasmonicApplications,bin2021ultra,Reshef2019MultiresonantMetasurfaces}. Unlike the LSPR, the SLR wavelength is defined by the periodicity of the positions of the nanoparticles in a periodic lattice, and this collective response has a relatively high $Q$-factor. For example, our recent work has experimentally demonstrated a $Q$-factor of  2340 around an operating wavelength of 1550~nm\cite{bin2021ultra}. Such an unprecedentedly high $Q$-factor in plasmonic nanostructures opens up a new door for designing high-$Q$-factor optical nano-devices.

In addition to improving the $Q$-factor of a lattice response, it is also highly desirable to be able to freely choose the resonance wavelengths of the metasurface\cite{Molesky2018InverseNanophotonics, Lassaline2020OpticalSurfaces}. The reason is simple: carefully tailoring multiple resonances in a metasurface allows one to customize a metasurface to specific applications, such as frequency comb generation\cite{Okawachi2011Octave-spanningChip, Kippenberg2011Microresonator-basedCombs}, ultra-sensitive bio-sensing\cite{Menezes2010Large-areaPhotovoltaics}, fluorescence enhancements\cite{Vecchi2009ShapingNanoantennas}. In metasurface systems, several methods have been studied to generate and tailor multiple lattice resonances, such as utilizing multiple materials\cite{Hsu2014TransparentScattering}, adding cladding layers to induce Fabry–Pérot SLR modes\cite{Reshef2019MultiresonantMetasurfaces}, removing nanoparticles periodically \cite{Zundel2021LatticeNanoparticles}, using all-dielectric Mie-resonance metasurfaces\cite{Kuznetsov2016OpticallyNanostructures}, and superlattice arrays \cite{Wang2015SuperlatticeArrays}. Each of these methods can support multiple resonance modes; however, they are all limited in the spectral position or number of these resonances. No method yet exists that can define any number of resonances in arbitrary locations in a metasurface. In this paper, we tackle this issue by exploiting a degree of freedom that has gone largely unexplored to date: by breaking the lattice periodicity, we show that we can simultaneously introduce multiple new resonances to a single metasurface. With this knowledge, we introduce and experimentally demonstrate the concept of Fourier Lattice Resonances (FLRs), which are a natural generalization of SLRs that can support multiple lattice resonances at desired operation wavelengths.

\section{Interpretation of Fourier Lattice Resonances (FLRs)}

For conceptual clarity, we begin by exploring how perturbations in translational symmetry introduce resonances. This demonstration helps interpret the physical origin of FLRs. Consider a metasurface consisting of rectangular gold nanoparticles with a periodic lattice arrangement. The lattice is embedded in a homogeneous background of silica glass ($n\,{\sim}\,1.45$), with  particle dimensions of $L_x = 110$~nm and $L_y = 180$~nm and periodicities of $P_x = 500$~nm and  $P_y = 1063$~nm (see the inset in Fig.~1(i)). The predicted transmission spectrum for an $x$-polarized beam is calculated by the finite-difference time-domain (FDTD) method and is shown in Fig.~1(i). The LSPR mode of the metasurface occurs at an operating wavelength of 830~nm, and the SLR is at 1550~nm ($\lambda_\mathrm{SLR} \approx nP_y$). Next, we break the symmetry of the rectangular lattice by displacing every second row of particles by a distance $\sigma= P_y/3$ (see the inset in Fig.~1(ii)). The calculated transmission profile of the metasurface is altered by this modulation (Fig.~1(ii)): a new collective resonance arises at 1040~nm. The origin of this resonance is clear when looking at Fig.~1(iii), where we show a rectangular lattice of periodicity ${P'}_y=2P_y/3$; the altered lattice in Fig.~1(ii) resembles a sparser version of Fig.~1(iii). The new resonance is, therefore, just an emergent SLR corresponding to this new periodicity, with a new resonant wavelength $\lambda'_\mathrm{SLR}=~n{P'}_y$. Figure~1(iii) shows the corresponding transmission profile with ${P'}_y=~2P_y/3$. Here we observe that the SLR with period $2P_y/3$ arises where the emergent SLR appears; hence the altered lattice arrangement can be viewed as combinations between two different periodic lattice arrangements. 

\begin{figure*}[ht!]
\includegraphics[width=\textwidth]{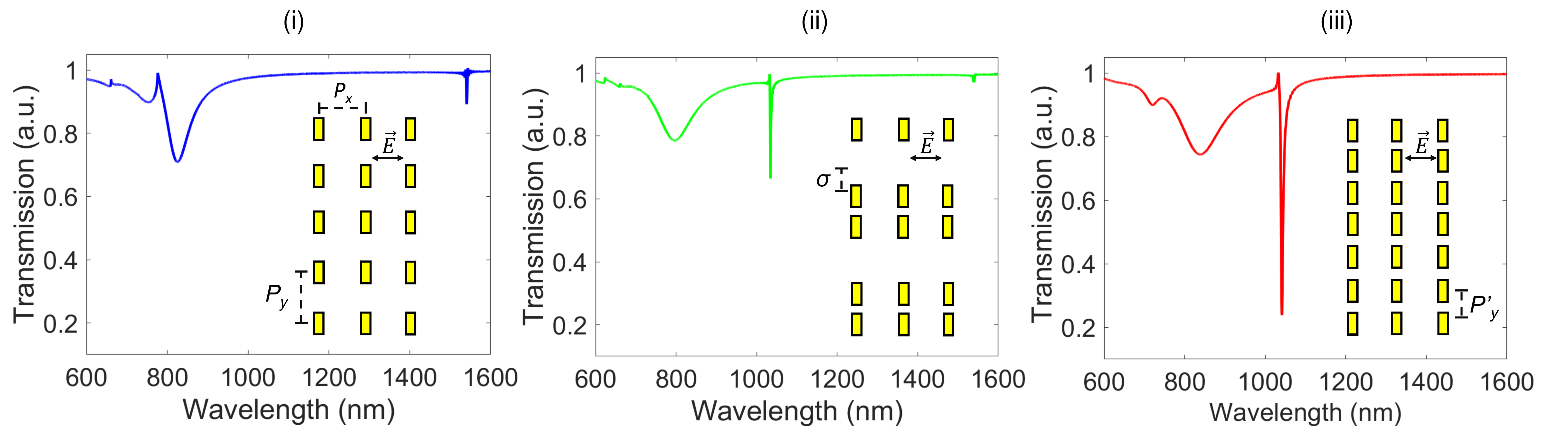}
    \caption{{\bf Emergent Surface Lattice Resonance modes. }{ The transmission profiles of three different lattice arrangements (i), (ii), and (iii) with their corresponding schematic drawings (see the inset figures). These lattice arrangements are made of rectangular nanoparticles, where $P_y$ or  ${P'}_y$ and $P_x$ are the lattice periodicity in the \emph{y}-axis and \emph{x}-axis, respectively. The incident electric field is polarized along the \emph{x}-axis. (i) and (iii) are periodic arrays with lattice constants of $P_y$ and ${P'}_y$, respectively. For the modulated lattice arrangement (ii), every second row of particles in the array in (i) is translated by spacing $\sigma = P_y/3$. Such system resembles the array of (iii) but with vacancies. }}
    \label{fig:final_image_1.pdf}
\end{figure*}

The near-field profile of the metasurface can provide additional clues for understanding this emergent SLR mode. Figure~2a illustrates the $x$-polarized electric field of all three lattices shown in Fig.~1. If we compare the electric field profile in Fig.~2a~(ii) with that of another perfect lattice arrangement with a periodicity of $2P_y/3$ at a wavelength of 1040~nm (Fig.~2a~(iii)), the same SLR modes are excited at the same wavelength. Hence, the emergent SLR can be regarded as an SLR mode with a periodicity of 2$P_y$/3, but missing particles. Additionally, the emergent SLR is not a diffraction order since the strength of the electric field confinements of the emergent SLR (Fig.~2b~(ii)) is comparable to both perfect lattice arrangements (Fig.~1(i) and Fig.~1(iii)).

\begin{figure*}[ht!]
    \includegraphics[width=\textwidth]{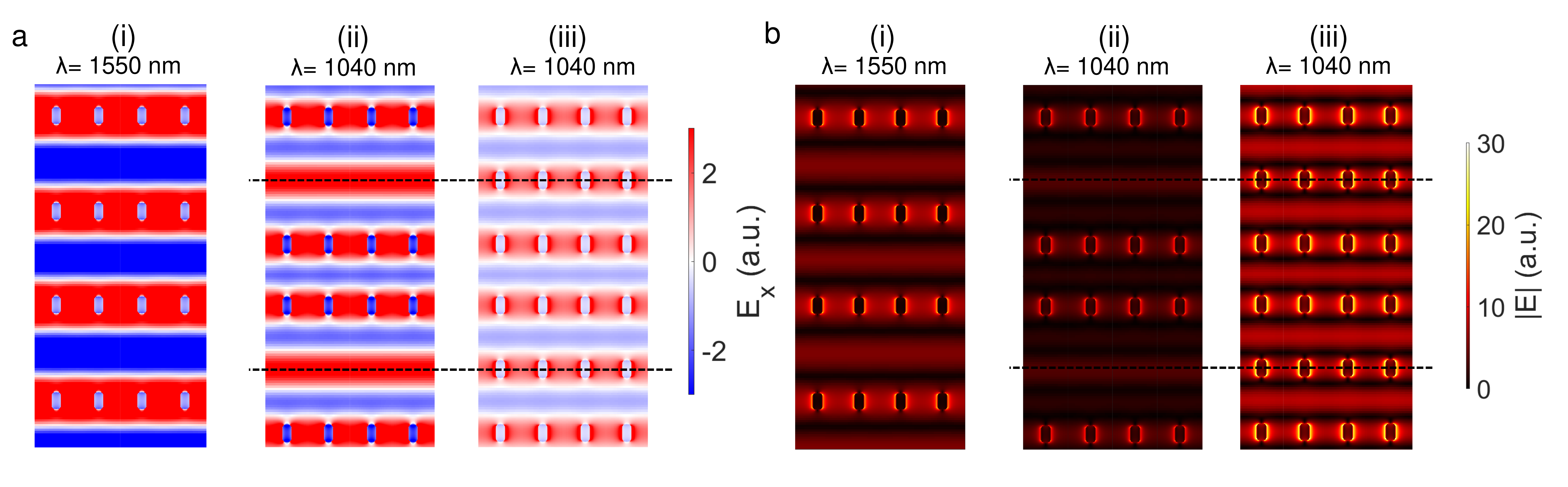}
    \caption{{\bf Near-field profiles of the lattice modes. }{\bf a) }{ The normalized $x$-polarized electric field profiles of lattice arrangements from Fig.~1, their respective SLR resonances of 1550 nm for (i) and 1040nm for (ii) and (iii). The vacancies in every second row of the unit cell (ii) are aligned with every second row of the particles of the unit cell in (iii)  (the black dashed line). the electric fields in (ii)  resemble those of (iii), indicating that a similar mode is being excited. }{\bf b) }{ The electric field amplitude profiles of the lattice arrangements, showing the field enhancement at the resonances. These results indicate that the emergent SLR is not a diffraction order since the field enhancement is within the same order of magnitude as the periodic lattices ((i) and (iii)).} }
    \label{fig:final_image_1_s2_1.pdf}
\end{figure*}

To observe the emergence of this SLR mode, we perform a series of FDTD simulations as a function of $\sigma$. We organize the  simulation results in Fig.~3a. As shown in this figure, for $\sigma$ = 0~nm, the emergent SLR does not appear at 1040~nm since the lattice arrangement is still rectangular. As $\sigma$ gradually increases, the periodic SLR mode at 1550~nm starts to vanish due to the introduction of lattice disorder, and the emergent SLR mode begins to appear when $\sigma$ is changed from 80~nm to 640~nm at a wavelength of 1040~nm. The most pronounced resonance occurs when $\sigma \approx 354$~nm since this corresponds to approximately a period of $2P_y/3$. 

We employ the lattice sum approach (LSA)\cite{Draine1994Discrete-DipoleCalculations, Zou2004SilverLineshapes} to gain further insight into these emergent modes. In this model, the effective polarizability induced in a single particle can be calculated by taking into account the change in the local field experienced by the particle due to the rescattered fields of the surrounding neighbour particles in the lattice. One of the main advantages of this method is that the computation time is much faster compared to a standard FDTD simulation with a superlattice element with periodic boundary conditions \cite{Reshef2019MultiresonantMetasurfaces, bin2021ultra}. On a standard desktop computer, the LSA simulations of a $150\times150\,\mu$m$^2$ lattice for the 2790 antenna placements shown in Fig.~3a took $\sim$1 minute for the entire spectral sweep (600-1600~nm), whereas the FDTD simulations of two-particles superlattice with periodic boundary condition took $\sim$1 hour. Clearly FDTD calculation of full complete metasurface is computationally prohibitive. In this paper, we modify the lattice sum by including the effect of $\sigma$ with a lattice size of $150\times150\,\mu$m$^2$, and perform the corresponding calculations concerning $\sigma$, as illustrated in Fig.~3b (see also Methods). By comparing this result with that shown in Fig.~3a, we demonstrate that the transmission spectra obtained by LSA calculations are in excellent agreement with FDTD simulations in terms of predicting the resonance wavelengths for the LSPR, periodic SLR, and emergent SLR.

\begin{figure*}[ht!]
\includegraphics[width=\textwidth]{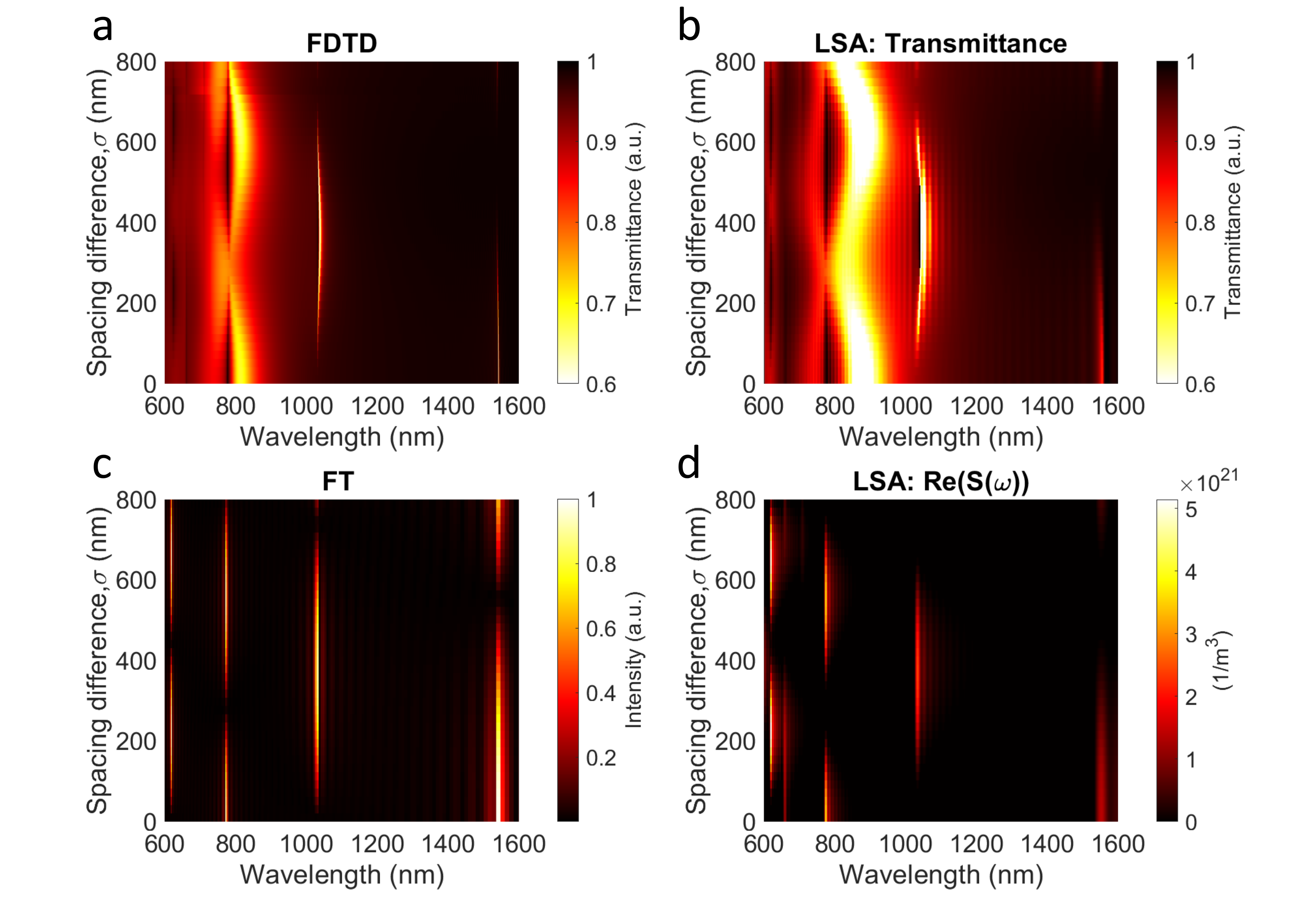}
    \caption{{\bf The evolution of the emergent SLR mode.} {\bf a) }{ The  FDTD-calculated transmission profiles with respects to the spacing difference $\sigma$. } {\bf b) }{ LSA-simulated transmission spectra plotted as a function of $\sigma$. The calculated LSPR and SLR modes are largely consistent with the calculations in (a). } {\bf c)}{ The FT-calculated transmission spectra with varying $\sigma$. }{\bf d) }{ The LSA-simulated real part of lattice sum function, Re$(S(\omega))$, plotted as a function of $\sigma$. The spectral features are consistent with the FT calculations in (c). } }
    \label{fig:Figure2_Data_Calculations_FFT.JPG}
\end{figure*}

The two SLRs in our example originate from the various periodicities present in the lattice ($P_y$ and $2P_y/3$). This suggests that taking a Fourier Transform (FT) of the particle locations may reveal the positions of the resonances. To demonstrate this result, we generate a function, $f(x)$, with a series of Gaussian profiles with a spacing corresponding to the periodicity of the actual lattice, $P_y$, between each profile. This function then mimics the actual lattice (see Method). We show this explicitly by calculating a series of FTs of $f(x)$ with the same corresponding range of $\sigma$ (Fig.~3c). On the other hand, figure~3d shows the LSA-simulated real part of lattice sum function, Re$(S(\omega))$, as a function of $\sigma$ (see equation (2)). These calculations provide only the lattice resonances; hence a direct comparison with FT calculations can be achieved. Therefore, we observed that the power spectrum (Fig.~3c) in the FT calculation has peaks in the exact wavelengths where the Re$(S(\omega))$ calculation features the lattice resonances in Fig.~3d. This agreement indicates that these new SLRs resonances can be predicted with a simple FT of the perturbed lattices; we call this predictive design approach the Fourier lattice resonance (FLR) method.

\section{Experimental Verifications of FLRs}

To experimentally verify the FLR model, we fabricated two sets of metasurfaces consisting of rectangular nanoparticles. In Fig.~4a, the images of the devices are presented; each sub-figure in Fig.~4a represents the actual device with different values of $\sigma$, and with a $P_y$ of 1063~nm. This set of metasurfaces has nanoparticle dimensions of $L_x$ = 110~nm and $L_y$ = 180~nm, and lattice sizes of $150\times\,150\,\mu$m$^2$. The calculated transmission spectra for these devices are shown in Fig.~4b and are in good agreement with the experimental data in Fig.~4c. The experimental setup and procedure are described in the Methods section. As predicted, the periodic SLR mode at 1040~nm vanishes as $\sigma$ increases. Furthermore, the emergent SLR starts arising when $\sigma$ = 88~nm, and is most prominent when $\sigma = 354$~nm, which agrees with the LSA and the FDTD simulations, supporting our theoretical interpretation.  Notably, the $Q$-factors of this set of devices range from 80 to 90, and the sharpest emergent SLR dip has a $Q$-factor of 90 with $P_y = 1063$~nm and $\sigma = 354$~nm. 

Our model also works with different values of $P_y$. Here, we fabricated another set of metasurfaces with $P_y = 1598$~nm (Fig.~4d) and the same nanoparticle sizes as the first set of the devices. Similarly to the first set of metasurfaces, the simulated transmission profiles of the second set of metasurfaces ($P_y = 1598$~nm) in Fig.~4e are also in good agreement with the experimental data (Fig.~4f), where the emergent SLR mode for this system is predicted to occur at 1550~nm and $\sigma = 354$~nm. Despite the extinction coefficients of the resonances in these metasurfaces being smaller than the first set, the $Q$-factors of these devices are relatively higher. The $Q$-factors range from 170 to 220, where the lattice resonance with the highest $Q$-factor appears at 1150~nm and $\sigma=~532$~nm. In the first set of devices, the spectral distance between the SLR at 1550~nm and the second SLR in our spectrum was 504~nm, in the second set, the distance is only 381~nm. This hints at the flexibility of modulated lattices to place SLRs at arbitrary spectral locations.

We have just shown that the LSA succeeds at describing the transmission spectra of a modified lattice arrangement. However, a discrepancy arises in the $Q$-factor predicted by this method when compared to FDTD simulations and the experimental results, which we attribute to the difference in array size \cite{Rodriguez2012CollectiveMatters,Zundel2018Finite-sizeNanostructures}. Because the nanoparticle density in the second set of metasurfaces is smaller although the array size is the same, the extinction ratio and $Q$-factor are lower than in the first set, making the second set of metasurfaces harder to characterize.

\begin{figure*}[tb]
    \includegraphics[width=\textwidth]{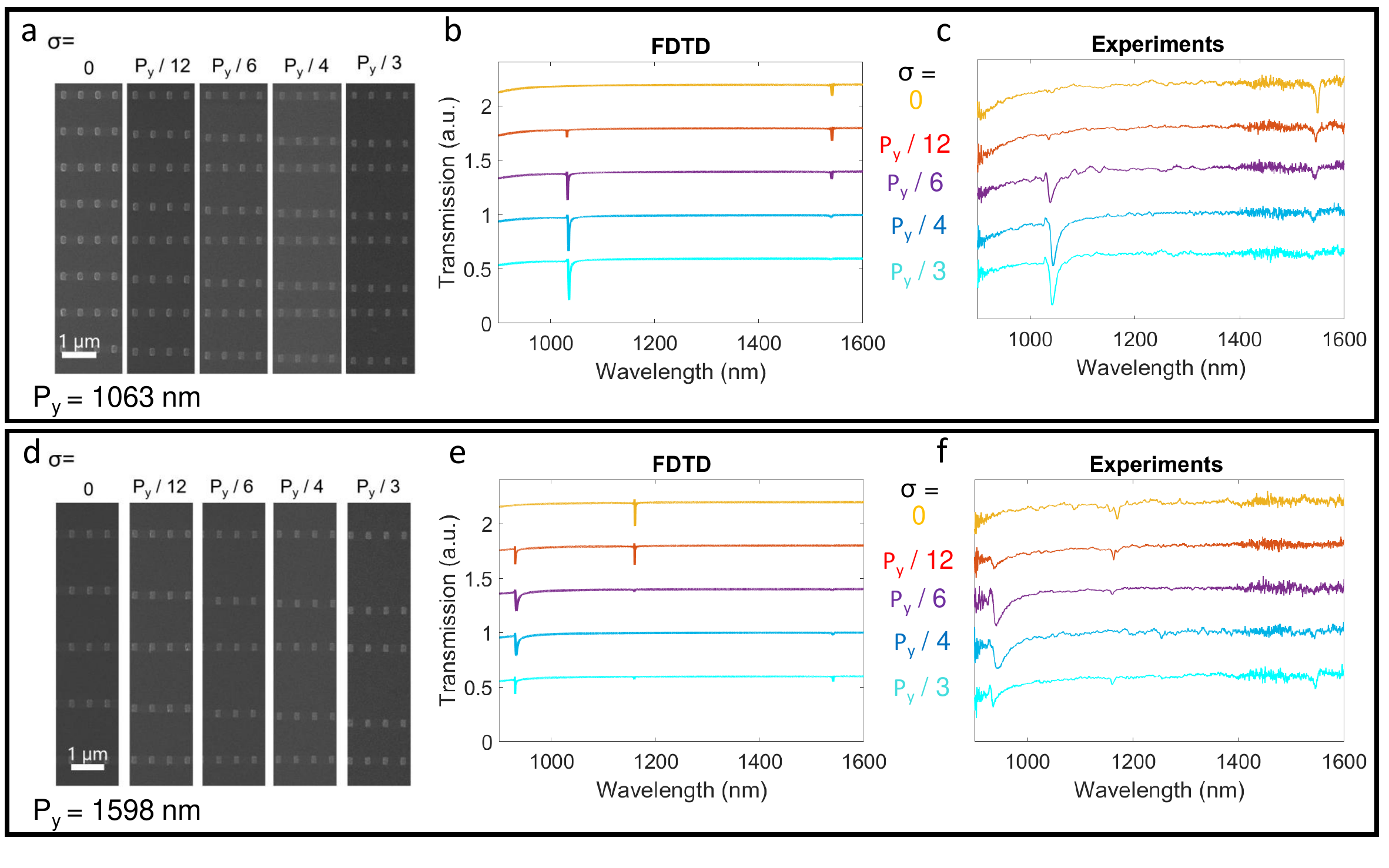}
    \caption{{\bf Experimental observation of emergent SLR modes. }{\bf a) and d) }{Helium ion microscopic images of the fabricated metasurfaces prior to cladding deposition. Each sub-figure corresponds to an image with a different $\sigma$. For (a), all metasurfaces have $P_y = 1063$~nm, whereas for the metasurfaces in (d), $P_y = 1598$~nm. } {\bf b) and e) }{ The simulated transmittance of the corresponding metasurfaces with (b) $P_y = 1063$~nm and (e) $P_y = 1598$~nm.} {\bf c) and f) }{The measured transmission spectra of the metasurfaces with (c) $P_y = 1063$~nm and (f) $P_y = 1598$~nm.  }  }
    \label{fig:final_image_3.pdf}
\end{figure*}

\section{Reverse-engineering of FLRs method}
Since the properties of the emergent SLRs can be predicted by using the FLR method, in principle, the process can be reverse-engineered. One could begin with the desired transmission profile, then use the FT of that spectrum to directly engineer the nanoparticle placement to generate the desired spectrum. To demonstrate this idea, we first select a set of desired resonance wavelengths and create a virtual spectrum (Fig.~5a), where we are deliberately placing two SLRs at 1150~nm and 1220~nm. We also include replicas of these two desired resonances represented as integer Nth harmonics in this virtual spectrum, because that there will naturally appear higher-order diffraction terms at $\lambda_\mathrm{SLR}$/2, $\lambda_\mathrm{SLR}$/3, $\emph{etc}$ for any lattice resonance. Thus, the FT-simulated transmission profile of a lattice must reproduce these corresponding features. We then take the inverse FT (IFT) of the virtual spectrum (Fig.~5a) and obtain the corresponding particle locations by using a peak-finder function (Fig.~5b). We consider the minimum spacing between nanoparticles to avoid overlapping particles while performing the calculations; we then set the minimum spacing of the particles to 250~nm. We also truncate some particles ($\sim$1-20) at the end of the array to a maximum desired array size. Finally, we calculate the transmission spectra of the IFT-generated lattice arrangement via LSA. We observe that the corresponding SLR modes for each set are induced at the desired wavelengths (Fig.~5d).

\begin{figure*}[htb!]
    \includegraphics[width=\textwidth]{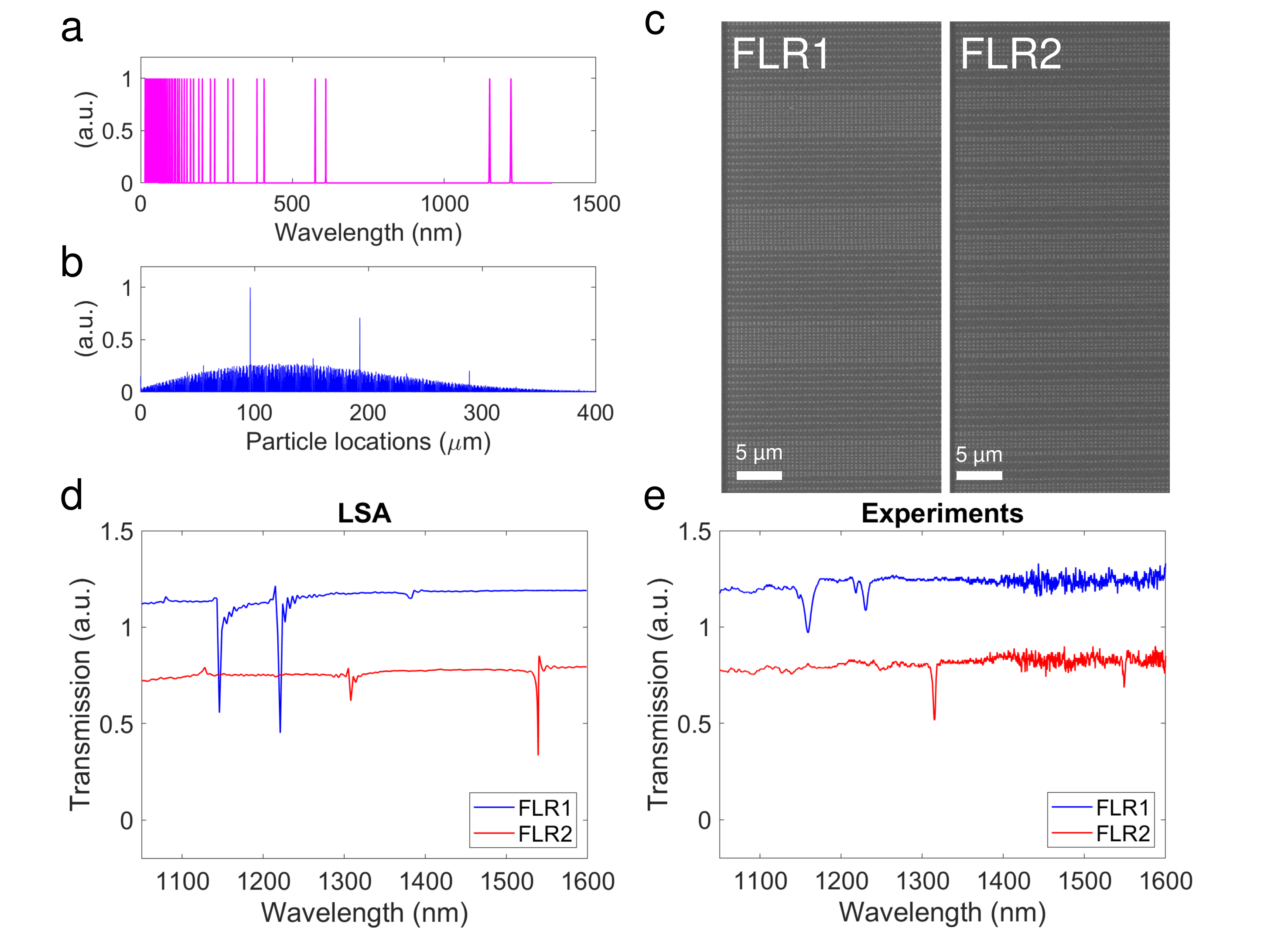}
    \caption{{\bf The FLR metasurface. }{\bf a) }{The generated virtual spectrum. We created a spectrum with resonance wavelengths of 1150~nm and 1220~nm, including their corresponding harmonics since each lattice resonance will naturally generate higher-order diffraction terms. The vertical axis of the figure is the virtual beam intensity, which we set to unity. }{\bf b)}{ The IFT of the virtual spectrum. The $x$-axis of the diagram indicates the corresponding particle locations for the desired wavelength combination. The vertical axis of the figure is IFT power spectrum intensity; we employ a peak-finder function with a small threshold (0.015) to determine the particle locations. The superlattice for FLR1 is non-repeating over $\sim$ 20 antenna in $y$-axis (here equivalent to $\sim$6 $\mu$m$^2$). Making FDTD simulation of this device prohibitive, but still possible with LSA.} {\bf c) }{SEM images of subsets of the lattices for FLR1 and FLR2 metasurfaces. }{\bf d) }{Calculated transmission spectra of the FLR metasurfaces using LSA method. The SLRs arise at the desired wavelengths.} {\bf e) } {The measured transmission spectra of the corresponding FLR metasurfaces.}  }
    \label{fig:final_image_4.pdf}
\end{figure*}

\section{Experimental demonstration of reverse-engineering capabilities of FLRs}

To experimentally demonstrate the reverse-engineering capabilities of the FLR method, we fabricate two devices with different sets of resonance wavelengths, one with resonances at $\lambda =$~1150~nm and 1220~nm (FLR1) as predicted here and another, equivalent lattice prepared for resonances at $\lambda =$~1310~nm and 1550~nm (FLR2), where the array size is $100\times\,400\,\mu$m$^2$ for both devices. Figure~5c shows images of subsets of the lattices for the FLR1 and FLR2 metasurfaces. After performing the measurements on these devices, we compare the LSA simulations (Fig.~5d) with the experimental data (Fig.~5e). The predicted resonance wavelengths from the LSA simulations are in good agreement with the experimental results of both devices. This agreement suggests that this method can be used to design resonance wavelengths flexibly. The $Q$-factors of the lattice resonances in the FLR1 metasurface are 130 at $\lambda =$~1150~nm and 200 at 1220~nm, and the $Q$-factors for the FLR2 metasurface are 400 at $\lambda =$~1310~nm and 800 at 1550~nm. Fabricating the same device with a larger array size would likely increase the depth and $Q$-factor of the resonance\cite{bin2021ultra, Zundel2018Finite-sizeNanostructures, Reshef2019MultiresonantMetasurfaces}.

The highest $Q$-factor of these FLRs is 800 (FLR2, at a wavelength of 1550~nm), which is two orders of magnitude higher than those associated with the LSPRs of the individual nanoparticles and an order of magnitude higher than that of Fabry–Pérot SLR modes \cite{Reshef2019MultiresonantMetasurfaces}. With such high $Q$-factors and the strong optical nonlinearity of metals\cite{bin2020hyperpolarizability}, the FLR metasurfaces could find applications in nonlinear processes, such as optical spontaneous parametric down-conversion (SPDC) and sum-frequency generation (SFG) \cite{Santiago-Cruz2021PhotonMetasurfaces, boyd2020nonlinear} because one can place the resonant wavelengths at the designated SPDC or SFG idler and signal wavelengths. Hence, significant SPDC or SFG efficiency enhancements can be achieved. 

To consider how existing approaches could be complemented by our FLR method, we consider a few recent, key examples from the literature: a recent work showed that a single dielectic nanoparticle could be engineered to have multiple resonances, some of which showed $Q$-factors $\sim$190\cite{Koshelev2020SubwavelengthNanophotonics}. This required careful simulation of the nanoparticle dimensions and precise fabrication. By comparison our FLR method could likely design a metasurface with similar multiple resonances from the spectrum itself. Obviously the nature of the resonators themselves are very different, but this exemplifies the ease of our design approach. Other related attempts in generating multiple resonances such as inducing Fabry–Pérot SLRs using a cladding \cite{Reshef2019MultiresonantMetasurfaces}, and using orthogonal polarizations\cite{Huttunen2019EfficientArrays, binalam2021crosspolarized} in the lattice are often restricted to resonances that are periodic in frequency. Our method is not restricted to periodic resonances, and it only assumes dipole effects of the particles, leading to more straightforward computations. 

Designing the FLR metasurface using standard FDTD simulations would be exceptionally challenging because the lack of symmetry in the lattice eliminates the possibility of simulating a small supercell, and the computational time increases with the size of the lattice arrangement\cite{10.1007/978-94-024-0850-8_25}. In contrast, the computational time required by the FLR method is reasonable (typically within 3 minutes), necessitating the computation of only a single FT per lattice. 

Other recent work has computed FTs to determine the 3D structure of a surface for applications in optical wavefront shaping; thus, fabricating such structure requires more sophisticated techniques\cite{Lassaline2020OpticalSurfaces, Mamin1992ThermomechanicalTip}. By contrast, our work aims to find a desired transmission spectrum, and our method only utilizes standard electron-beam lithography techniques instead of 3D structures, which leads to a more straightforward fabrication. 

Other related work has extensively demonstrated designing multi-layer thin film stacks using FTs for applications in optical filtering \cite{Dobrowolski1978OpticalTransforms}. Noting the similarity of our work to the study of Ref. \cite{Dobrowolski1978OpticalTransforms}, we believe that our methodology could be further extended to 3D lattices by combining our design method with the developed analytical method.

Future work on this platform could look into modifying the nanoparticle geometries to tune the width and depth of the individual resonances. Furthermore,  we only took the position of the particle locations and ignored the magnitude of the peaks in the power spectrum when performing the IFTs. Some information is encoded in these magnitudes on the scattering amplitude of individual nanoparticles in the array. This information could perhaps be incorporated in future metasurfaces to modulate the individual extinction ratio of each resonance.

The $Q$-factor of the FLRs can be increased further by using dielectric particles. However, in those cases, higher-order multipoles would need to be taken into consideration\cite{Koshelev2020SubwavelengthNanophotonics}. In this work, we did not modify the arrangement of the particles in the $x$-direction. We note that such modifications should give rise to a different set of resonances that could be excited by using orthogonal polarization, i.e. $y$-polarized light\cite{Huttunen2019EfficientArrays}. One possible application would be to modify the LSA calculations to generate orbital angular momentum (OAM) modes by introducing particle size disorders or different particle geometries.

\section{Conclusion}

In summary, we have theoretically proposed and experimentally demonstrated the concept of FLRs. The fabricated devices in this paper enable SLRs at multiple simultaneous wavelengths, and the experimental results are in good agreement with our FLR calculations. Additionally, these devices require only a standard fabrication procedure, and they can be easily expanded upon to generate multiple resonances that can be placed arbitrarily. From a broader perspective, this method is possible because we exploit a previously ignored degree of freedom by breaking the lattice periodicity, and this modelling is only possible thanks to large-scale modelling techniques like LSA. We anticipate that this novel design method will become significant in realizing resonant metasurfaces for many optical applications.

\section*{Methods}

\paragraph{Simulation.}  A commercial three-dimensional finite-difference time-domain (FDTD) software package is utilized to perform full-wave simulations for single rectangular nanoparticles, periodic metasurfaces, and modulated metasurfaces. For the single nanoparticle simulations, the total-field scattered-field (TFSF) method is used to extract the single particle polarizability in all three dimensions: $\alpha_{xx}$, $\alpha_{yy}$, and $\alpha_{zz}$ \cite{Alaee2017TheoryAbsorbers}. For the periodic and modulated metasurfaces, a single unit cell was simulated using periodic boundary conditions in the in-plane dimensions and perfectly matched layers in the out-of-plane dimension. All three cases were modeled using fully dispersive optical material properties for silica and for gold\cite{P.B.JohnsonandR.W.Christy1972OpticalMetals}. Minimal artificial absorption (Im\emph{$(n)$} $\approx$ $10^{-4}$) was added to the background medium to reduce numerical divergences.

The lattice sum approach (LSA) is a simplified version of the
discrete-dipole approximation (DDA) method\cite{Draine1994Discrete-DipoleCalculations, Zou2004SilverLineshapes}. The main difference between these two methods is that the LSA assumes that the dipole moments of all interacting nanoparticles are identical. Thus, the dipole moment of any nanoparticle can be simplified as

\begin{equation}
    \mathbf{p}=\frac{\epsilon_{0} \alpha(\omega) \mathbf{E}_{\text {inc }}}{1-\epsilon_{0} \alpha(\omega) S(\omega)} \equiv \epsilon_{0} \alpha^{*}(\omega) \mathbf{E}_{\text {inc },}
\end{equation}
\noindent
where the effect of inter-particle coupling is incorporated in the lattice sum $S(\omega)$, $\alpha^*(\omega)$ is the effective polarizability, and $\alpha(\omega)$ is the single nanoparticle polarizability obtained from the TFSF method. The $S(\omega)$ can be written as

\begin{equation}
    S(\omega)=\sum_{j=1}^{N} \frac{\exp \left(\mathrm{i} k r_{j}\right)}{\epsilon_{0} r_{j}}\left[k^{2}+\frac{\left(1-\mathrm{i} k r_{j}\right)\left(3 \cos ^{2} \theta_{j}-1\right)}{r_{j}^{2}}\right]
\end{equation}
\noindent
where \emph{$r_j$} is the spacing to the \emph{j}th dipole, \emph{k} is the wavenumber, and \emph{$\theta_j$} is the angle between \emph{$r_j$}. For a periodic metasurface, \emph{$r_j$} is just the periodicity  \emph{$P_v$} multiplied by \emph{j} and the indices \emph{v} indicates the \emph{x}- or \emph{y}-axis. 

The function $f(x)$ that mimics the lattice arrangement is considered by producing a series of Gaussian profiles. Each Gaussian profile represents a nanoparticle, and the Gaussian peak intensity is set to 1. We set the bandwidths of these Gaussian profiles to be significantly smaller than the spacing between the particles  (bandwidth-to-spacing ratio is $\sim$100); thus, we can assume that the particles act like point-like dipoles under the applied incident field, similar to the assumption made in the LSA model. Next, the spacing between each Gaussian profile is $P_y$ in the periodic case. When we considered a modulated lattice arrangement, the position of every second nanoparticle was shifted by $\sigma$. Then, we performed FT of the function with respects to $y$-axis, and we obtained the density plot in Fig.~3c.

\paragraph{Fabrication.} The metasurfaces are fabricated using a standard metal lift-off process and a positive-tone resist bilayer. We started with a fused silica substrate and defined the pattern using electron-beam lithography with the help of a commercial conductive polymer. The mask was designed using shape correction proximity error correction\cite{Schulz2015QuantifyingInvited} to compensate for corner rounding. Following the development, gold was deposited using thermal evaporation. The final silica cladding layer was deposited using sputtering. The backside of the silica substrate was coated with an antireflective coating to minimize substrate-related etalon fringes. 

\paragraph{Experimental setup.}  A collimated broadband supercontinuum laser source was used to illuminate the devices with a wavelength spectrum ranging from 470~nm to 2400~nm. We used a linear polarizing optical filter to control the light beam polarization. Then, the light beam was transmitted through the device and a lens imaged the sample onto the detector array. We placed a pinhole with a diameter of $100\,\mu$m in the image plane to choose the desired metasurface. After that, the remaining light beam was guided into  a multimode fiber with a diameter of $400\,\mu$m, which was coupled into an optical spectrum analyzer. Finally, we normalized the transmission spectrum of the light beam to a background trace of the substrate without gold nanostructures. 

\section*{Author contributions}
TL and OR conceived the basic idea for this work. TL and YV conducted the transmission measurements. TL, OR, and MSBA designed the experiment. RA and LC calculated single particle's polarizability. TL and OR carried out the simulations. TL, OR, YV, and JU analyzed the experimental results. RWB, JU, MJH, and KD supervised the research and the development of the manuscript. TL and OR wrote the first draft of the manuscript, and all authors subsequently took part in the revision process and approved the final copy of the manuscript.

\begin{acknowledgement}
We acknowledge the help of Sabaa Rashid with imaging. We thank Brian T. Sullivan and Graham Carlow for fruitful discussions. Fabrication in this work was performed in part at the Centre for Research in Photonics at the University of Ottawa (CRPuO). This research was undertaken thanks in part to funding from the Canada First Research Excellence Fund and the Canada Research Chairs program. LC acknowledges the support of the China Scholarship Council. MJH acknowledges the Flagship of Photonics Research and Innovation (PREIN) funded by the Academy of Finland (Grant No. 320165). RA acknowledges support from the Alexander von Humboldt Foundation through the Feodor Lynen Return Research Fellowship. We acknowledge the support of the Natural Sciences and Engineering Research Council of Canada (NSERC) [funding reference number RGPIN/2017-06880, RGPIN/2020-03989, 950-231657 and STPGP/521619-2018].
\end{acknowledgement}

\providecommand{\latin}[1]{#1}
\providecommand*\mcitethebibliography{\thebibliography}
\csname @ifundefined\endcsname{endmcitethebibliography}
  {\let\endmcitethebibliography\endthebibliography}{}


\end{document}